# Superposed Wave (s-Wave): Accelerating Photoacoustic Simulation

Jiadong Zhang†, Tengbo Lyu†, Changchun Yang, Yimeng Yang, Shanshan Guo,
Feng Gao and Fei Gao*, *Member, IEEE*

*Abstract*— Photoacoustic imaging develops very fast in recent years due to its superior performance in many preclinical and clinical applications. However, it is still in a developing stage, and a lot of experiments have to be performed in a simulation setting. To simulate photoacoustic imaging in a computer, k-Wave is currently the most popular MATLAB toolbox. Lots of research groups choose k-Wave toolbox to perform the forward projection process, which also can be described as forward model. However, by solving complex partial differential equation, k-Wave suffers a lot from computation time. To accelerate photoacoustic simulation, in this paper, we propose a straightforward simulation approach based on superposed Wave (s-Wave). Specifically, we treat the initial pressure distribution as a set of single pixels. Then by pre-obtaining a standard sensor data from single pixel, we can easily use loop and multiplication operators to change phase and amplitude of sensor data for given pixels. We use three different 2D samples and two 3D samples to test the time cost. The result of our proposed s-Wave method shows much less time consumption compared with k-wave. Especially in a sparse configuration in 3D, s-Wave is more than 2000 times faster than k-Wave, whiling getting nearly same sensor data.

*Clinical Relevance*—This work saves much time in photoacoustic forward model simulation, which makes photoacoustic research easier and more convenient.

## I. INTRODUCTION

Based on photoacoustic (PA) effect, PA imaging combines high contrast of optical imaging and deep penetration of ultrasonic imaging. Induced by a pulsed laser, biological tissue emits ultrasound signals followed by thermal elastic expansion, which can be detected by ultrasound transducers [1][2]. This procedure can be described as forward projection. With these ultrasound signals, we can reconstruct PA imaging with different reconstruction algorithms, which can be described as backward projection. Forward projection is a physical procedure of wave propagation. On the other hand, backward projection is an approximate to inverse procedure of forward projection to rebuild image from PA signal.

Appearance of k-Wave [3] makes forward projection simulation available in computer environment conveniently. As a specific PA simulation toolbox, k-Wave provides different choices for user to simulate PA effect and wave propagation in medium [4][5]. It does matters in three ways: (1) Predict possible experiment phenomenon before practical experiment. (2) Test hardware limitation to save cost (e.g. number of transducers). (3) Generate large amount of useful PA dataset for reconstruction algorithm design. It's recognized

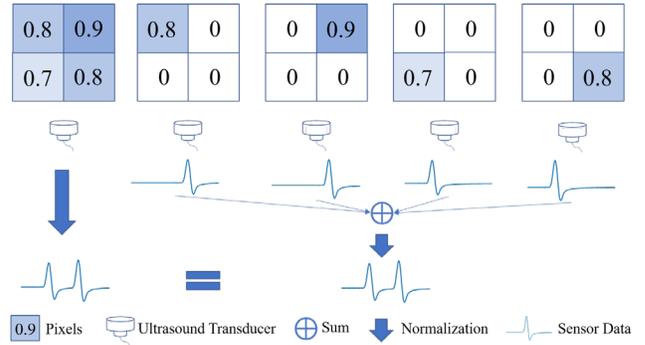

Fig 1. Illustration of sensor data induced by the whole part and every individual pixel in k-Wave simulation setting.

that k-Wave is the most popular PA simulation toolbox up to now.

While enjoining its convenience, k-wave simulation is time-consuming. The core of k-Wave is solving partial differential equation (Eq. (1)) in a numerical way [3].

$$\begin{aligned}(\partial_{tt} - c^2\Delta)p(x,t) &= 0\\ p(x,0) &= f(x)\\ \partial_t p(x,0) &= 0\end{aligned} \quad (1)$$

where $c$ is sound speed, $p(x,t)$ is acoustic pressure and $f(x)$ is the initial photoacoustic pressure distribution. Solving Eq. (1) means k-Wave has to compute every grid at any given time point. When grid size or sample rate increase, the computation time increases dramatically. So, how to find an easier and faster simulation substitution becomes an urgent challenge to be solved.

In this paper, we propose a novel simulation method (s-Wave) to get nearly same PA signal but with significantly less computation cost, compared with k-Wave. Considering superposition of wave [6][7], we use every single pixel of initial pressure distribution to generate specific PA signal and sum them up at the sensor's location, instead of treating it as propagating wave. Signal of different pixels mainly varies only in phase and amplitude, which can be computed with easy loop and multiplication operators. It avoids to solve complex Eq. (1), and saves much time, especially in high sample rate or large grid size conditions in sparse configuration.

We compared s-Wave with k-Wave in different conditions. While getting nearly same PA signal, we pay more attention to its computation time. The result shows that we can speed up the PA simulation procedure significantly in most conditions. Last, we give some possible practical applications of our method in future work.

†Those authors contributed equally to this work.
Jiadong Zhang, Tengbo Lyu, Changchun Yang, Yimeng Yang, Shanshan Guo, Feng Gao and Fei Gao are with the Hybrid Imaging System Laboratory, Shanghai Engineering Research Center of Intelligent Vision and Imaging, School of Information Science and Technology, ShanghaiTech University, Shanghai 201210, China (* corresponding author: gaofei@shanghaitech.edu.cn).

## II. METHOD

In this section, we will give detailed description about our proposed method, which is motivated by superposition of waves. Taking the 2D simulation as example, we hypothesize that there is relationship between the sensor data (PA signal) induced by every pixel's summation and conventional k-wave simulation of PA wave propagation. To validate this assumption, we use every pixel's signal in the whole initial pressure distribution to generate sensor data, which can be generalized in Fig. 1.

In Fig. 1, the left part illustrates the process of sensor data induced by conventional k-wave simulation. We use four pixels to generalize most situations. The right part illustrates sensor data induced by individual pixel. We only keep one pixel valid while setting other pixels to zero. By summing the individual sensor data up, we can find that the summated signal is almost as same as the sensor data generated by the conventional k-wave simulation after normalization. Besides, we also find that the individual sensor data's shape is similar with each other, they only vary in amplitude and phase. So, we propose to build received sensor data by simply delaying and summing standard sensor data induced by single pixel.

Taking 2D simulation as example, we can summarize our method as following four steps:

(1) According to medium's physical properties, sampling rate and initial pressure distribution, we can pre-obtain a standard sensor data $S$ with k-Wave [3], or other physical simulation software [8]. The standard sensor data $S$ features as follow: it is induced by a single pixel and its pixel value is one; without energy dissipation or attenuation; corresponding distance $d$ between pixel and sensor position is known.

(2) Change sensor data phase with loop operation for every pixel. Given a pixel position and a transducer position, we can get corresponding distance $d_p$. Then the sensor data of this given pixel can be obtained with phase operator $\tau$, which can be described as:

$$\tau(d_p, d) = \frac{(d_p - d)f}{v} \qquad (2)$$

where $f$ denotes sampling frequency, and $v$ denotes ultrasound wave propagation velocity in corresponding medium. When $\tau > 0$, it means $d_p > d$ and we have to right shift standard sensor data $S$ to get the right phase. Otherwise we have to left shift standard sensor data $S$.

(3) Change sensor data's amplitude with multiplication operation for every pixel. There are three factors that will influence sensor data's amplitude: pixel value; energy dissipation; attenuation. In our method, we consider a more ideal situation, so we ignore attenuation's influence in this work. Because we normalize initial pressure distribution

TABLE I.  PSEUDOCODE OF PROPOSED S-WAVE

**Input:** standard sensor data $S$; corresponding distance $d$; $n \times m$ pixels with {pixel value $p$ pixel position $(x_p, y_p)$ }; $w$ transducers with {transducer position $(x_w, y_w)$ }.

1. *for i=1, 2, ..., $w$ do*
2.     *for j=1, 2, ..., $n \times m$ do*
3.        $d_p^{(j)} = \sqrt{(x_p^{(j)} - x_w^{(j)})^2 + (y_p^{(j)} - y_w^{(j)})^2}$
4.        Compute $\tau(d_p^{(j)}, d)$ with equation (2)
5.        Compute $A(d_p^{(j)}; p)$ with equation (4)
6.        Compute $s_p^{(j)}$ with equation (5)
7.     end
8.     $s^{(i)} = \sum_{j=1}^{n \times m} s_p^{(j)}$
9. *end*

**Output:** sensor data $s^{(1)}, s^{(2)}, ..., s^{(w)}$

ranging from 0 to 1, it means that this pixel value $p$ cannot surpass 1. In addition, because amplitude of sensor data is proportional to pixel value, we can directly multiply $p$ to standard sensor data $S$ to change its amplitude. Besides, considering the influence by energy dissipation, we use a reciprocal of a power function to simulate this process:

$$g(d_p) = \frac{k}{d_p^2} \qquad (3)$$

where $k$ is a constant coefficient, which can be computed by standard sensor data $S$. So we can describe amplitude operator as:

$$A(d_p; p) = p \times g(d_p) = \frac{kp}{d_p^2} \qquad (4)$$

(4) We repeat (1)-(3) for every pixel in given ultrasound transducer position. The sensor data $s_p$ can be described as:

$$s_p = A(d_p; p) \cdot \langle s \mid \tau(d_p, d) \rangle \qquad (5)$$

where $\langle \cdot \mid \cdot \rangle$ denotes loop with loop operators. By summing up every sensor data induced by single pixel, we can get the final sensor data.

(5) For different transducer position, we repeat steps (2) - (4), then we can get the whole set of sensor data for any given transducer position.

The pseudocode of 2D simulation situation to make this procedure clearer is shown in Table 1.

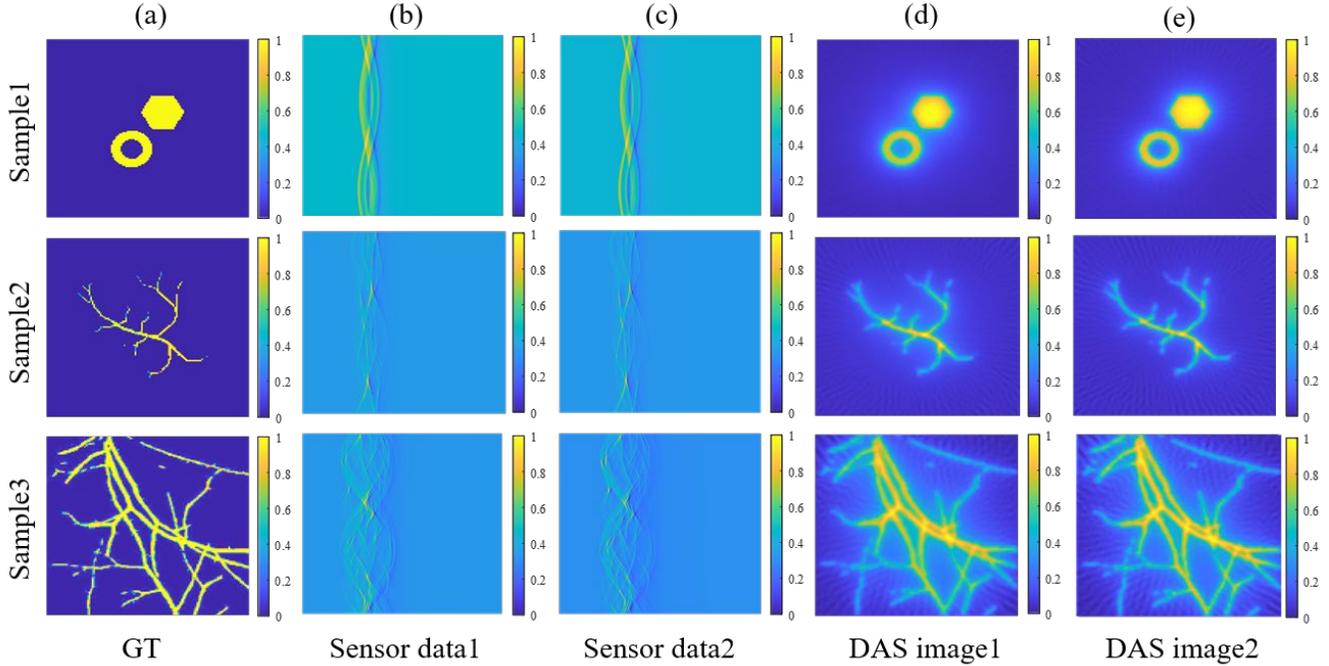

Fig 2. Sensor data and DAS images of different samples computed with k-Wave and s-Wave. (a) Initial pressure distribution (Ground Truth, GT). (b) Sensor data from k-Wave simulation. (c) Sensor data from s-Wave simulation. (d) Reconstruction of PA imaging from sensor data 1 with DAS algorithm. (e) Reconstruction of PA imaging from sensor data 2 with DAS algorithm.

## III. EXPERIMENTAL RESULTS

In order to prove the effectiveness of proposed method, we choose three different types of 2D samples and two 3D samples to perform the experiment. 2D samples are simple object, simple vessel network and complicated vessel network. The first two samples are standard numerical phantoms in k-Wave toolbox [3]. 3D samples are simple sphere object and skin vessel network [9].

First, we want to show that s-Wave can get nearly same result of 2D samples as k-Wave, which will be shown in section A below. Then we will compare time cost of 2D samples in different situations with k-wave simulation results in section B below. In section C, we extend s-Wave to 3D simulation situation. We simply compare results from s-Wave of two 3D samples with k-Wave simulation, in terms of both sensor data accuracy and time cost.

### A. Comparison of accuracy of 2D simulation

In Fig. 2, we show the imaging results of different types of samples with different simulation methods. In the same system setting, we use different 2D samples as initial pressure distribution (Fig. 2(a)). We use *Sensor data1* and *DAS image1* to denote the results from k-Wave simulation, and *Sensor data2* and *DAS image2* to denote the results from s-Wave. The second and third columns (Fig. 2(b) and 2(c)) illustrate sensor data, which are obtained from k-Wave simulation and s-Wave simulation. These two sets of sensor data are nearly same after normalization. In further, we reconstruct PA images with delay-and-sum (DAS) algorithms [10]. In order to show similarity of the *DAS image1* and *DAS image2*, we calculate their structure similarity (SSIM) index. SSIM ranges from 0 to 1. SSIM closer to 1 means two images are more alike. After threshold artifact removal, the imaging results achieve 95.17%, 91.50%, and 85.04% SSIM index for these three 2D samples.

### B. Comparison of time cost of 2D simulation

In this part, we pay more attention on time cost of the two methods. We analyze the factors, which may influence running time including pixel resolution of initial pressure distribution, sampling rate, number of transducers and transducer distance. The whole experimental design can be generalized in Table 2. In every experiment, we repeat 30 time and use averaged value as final time cost. All experiments are carried on Intel Core i5-9500E Processor with 8GB memory.

(1) Grid size (Experiment 1): Different pixel resolution of initial pressure distribution needs different grid size. In k-Wave simulation, it has to compute every grid, which induces tremendous time cost when the grid size increases. For s-Wave, we just have to compute most valuable pixels (pixel value that is larger than a fixed value) in step 2 and 3 in section 2. Overall, s-Wave works much faster than k-Wave for all the three types of 2D samples. For some specific situations, such as when the grid size of *sample 2* is 128, the improvement is more than 50 times. The quantitative results are summarized in Table 3.

(2) Sampling rate (Experiment 2): Higher sampling rate means that in the same time interval, we have to compute at more time points. It will both affect time cost of k-Wave and s-Wave. The consuming time has been summed up in Table 4. From the quantitative result, we can see that s-Wave is still much faster than k-Wave, especially for *sample 1* and *sample 2* with higher degree of sparsity. For some specific situations, such as when sampling rate of *sample 2* is 0.5 MHz, the improvement is more than 50 times.

(3) The number of transducers (Experiment 3): The number of transducers will not influence the time cost in k-Wave simulation, but will greatly affect the computation time of s-Wave. Because we have to repeat step (2)-(4) in section 2 for every transducer, it means more transducers will cause more time cost. But our method still works well in most time with less time consumption compared with k-wave simulation. The time cost is plotted in Fig. 3. The three sub-figures correspond to three samples, respectively. We can note that the time cost of s-Wave for *sample 1* and *sample 2* are still less than k-Wave, even if the number of transducers is up to 512. But for *sample 3*, s-Wave is faster when the number of transducers is less than 384. When more transducers are placed around the tissue, k-Wave will surpass s-Wave. Considering we don't need too many transducers in most situations, we believe s-Wave still can work well and faster than k-wave simulation.

(4) Transducer distance (Experiment 4): Theoretically, transducer distance will not affect time cost in ideal environment. But in k-Wave simulation, we have to place transducers in the grid, which means the longer the distance is, the larger grid size we need. So transducer distance will greatly affect time cost in k-Wave simulation. For s-Wave, the longer transducer distance means more time points, so we have to perform more loop operations. It slightly increases the time cost. For some specific situations, such as when transducer distance of sample 2 is 32.0 mm, the improvement is more than 90 times. The overall result is shown in Table 5.

*C. Comparison of 3D simulation*

We can easily extend s-Wave from 2D simulation to 3D simulation. We choose two different types of 3D samples: simple sphere object (*sample 4*) and complicated skin vessel network (*sample 5*) [9]. *Sample 4* has smaller grid size: $12 \times 24 \times 24$, and *sample 5* has larger grid size: $36 \times 128 \times 128$. Nine transducers are places under the samples. The nine transducers are placed in a $3 \times 3$ square flat way, the distance between the lower surface of samples and the plane of transducers is 12.8mm. Sampling rate is set as 0.5 MHz.

We show the results of k-Wave and s-Wave for these two 3D samples in Fig. 4. Fig. 4(a)-(b) illustrate the ground truth of samples (3D view), Fig. 4(c)-(f) show the sensor data obtained with k-Wave and s-Wave of the two samples. The two sets of sensor data from k-Wave and s-Wave show very high similarity with 99.82% and 99.87% SSIM index respectively.

With larger grid size in 3D compared with 2D simulation, the superiority of s-Wave is even more obvious. For *sample 4*, k-Wave costs 40.3575s. But s-Wave only needs 0.1794s, with 224 times faster. For *sample 5*, k-Wave costs 713.7793s. But s-Wave only needs 0.3479s, with 2051 times faster. These results in 3D simulation are exciting, because to simulate real situation, we always need to compute PA effect in 3D with much larger grid size.

TABLE II. EXPERIMENT DESIGN SUMMARY

|  | Grid Size | Sampling Rate(MHz) | Number of Transducers | Transducer Distance(mm) |
|---|---|---|---|---|
| Experiment 1 | 64/96/128/256 | 0.5 | 128 | 19.2 |
| Experiment 2 | 128 | 0.5/1/1.5/2 | 128 | 19.2 |
| Experiment 3 | 128 | 0.5 | 32~512(32) | 19.2 |
| Experiment 4 | 128 | 0.5 | 128 | 12.8/19.2/25.6/32.0 |

TABLE III. TIME COST (S) OF EXPERIMENT 1

|  | Sample 1 | | Sample 2 | | Sample 3 | |
|---|---|---|---|---|---|---|
|  | k-Wave | s-Wave | k-Wave | s-Wave | k-Wave | s-Wave |
| 64 | 1.8989 | 0.3675 | 1.7938 | 0.0161 | 1.8177 | 0.3675 |
| 96 | 4.6308 | 2.1207 | 4.6370 | 0.0961 | 4.6344 | 2.1207 |
| 128 | 37.3421 | 12.5687 | 38.0100 | 0.6925 | 37.4256 | 12.5687 |
| 256 | 419.0902 | 163.2513 | 413.4193 | 7.2325 | 418.9278 | 163.2513 |

TABLE IV. TIME COST (S) OF EXPERIMENT 2

|  | Sample 1 | | Sample 2 | | Sample 3 | |
|---|---|---|---|---|---|---|
|  | k-Wave | s-Wave | k-Wave | s-Wave | k-Wave | s-Wave |
| 0.5 MHz | 37.4123 | 1.4056 | 37.4157 | 0.6911 | 37.4002 | 12.6470 |
| 1 MHz | 75.6313 | 3.2377 | 74.8405 | 1.5671 | 74.5368 | 30.7605 |
| 1.5 MHz | 111.8738 | 5.0870 | 112.2771 | 2.4843 | 112.2051 | 49.2789 |
| 2 MHz | 149.4698 | 7.3897 | 149.1031 | 3.5535 | 149.0981 | 70.0846 |

TABLE V. TIME COST (S) OF EXPERIMENT 4

|  | Sample 1 | | Sample 2 | | Sample 3 | |
| --- | --- | --- | --- | --- | --- | --- |
|  | k-Wave | s-Wave | k-Wave | s-Wave | k-Wave | s-Wave |
| 12.8 mm | 14.8568 | 0.8090 | 14.8526 | 0.4109 | 14.8698 | 6.9036 |
| 19.2 mm | 37.2426 | 1.4355 | 37.2487 | 0.7054 | 37.3149 | 12.6137 |
| 25.6 mm | 71.8281 | 2.7269 | 71.8745 | 1.2012 | 71.8689 | 25.9916 |
| 32.0 mm | 135.8711 | 3.2867 | 136.2420 | 1.4403 | 136.1847 | 32.8496 |

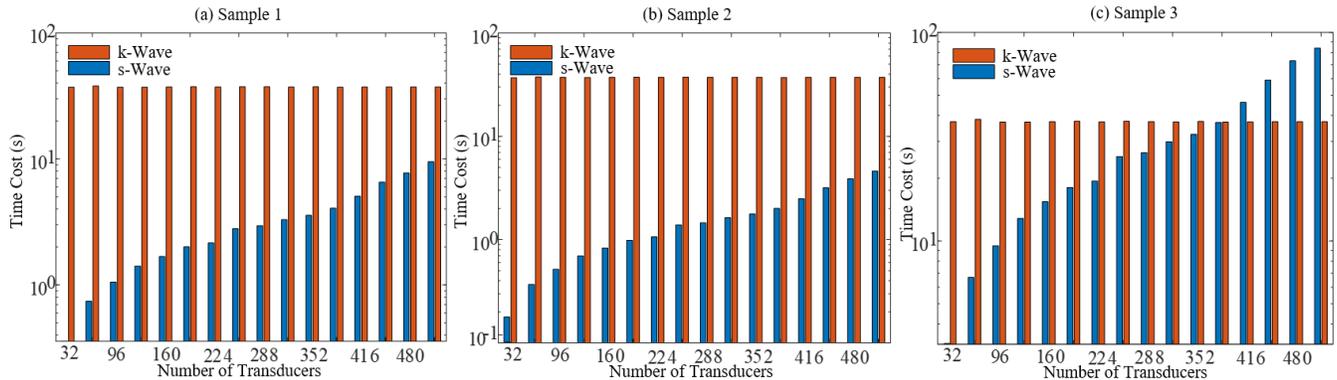

Fig. 3. Time cost of experiment 3. (a) (b) (c) subfigures illustrate the comparison of time cost with different transducer number (from 32 to 512) for three samples.

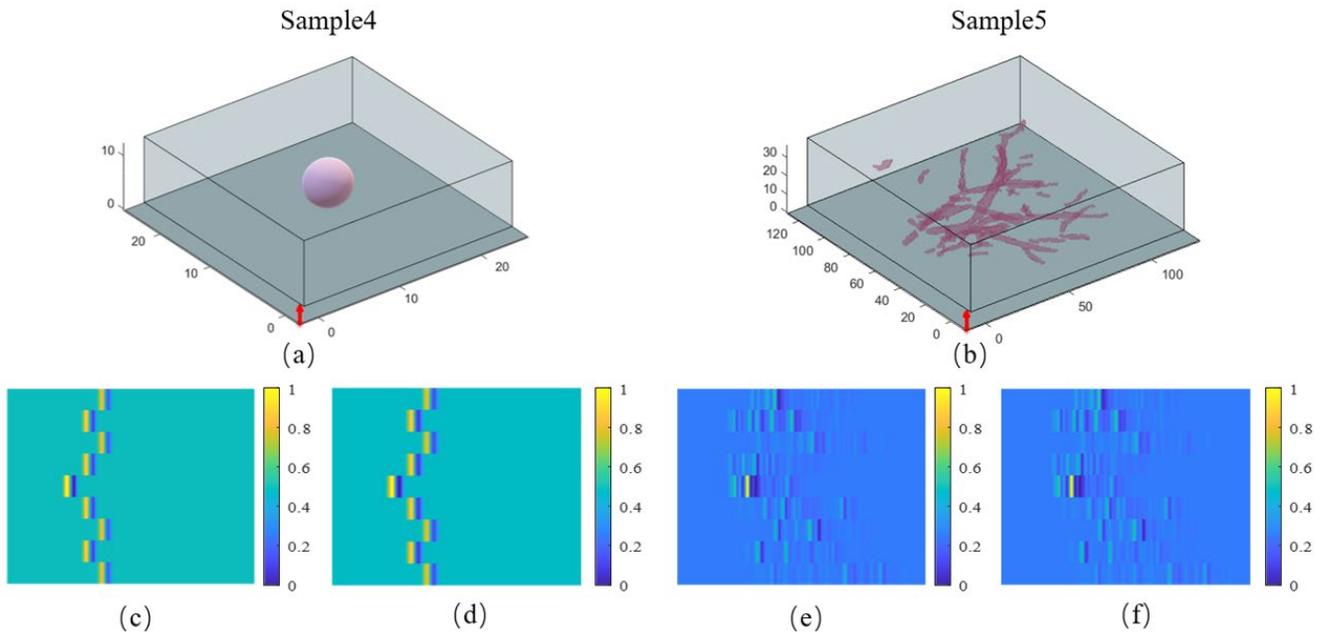

Fig. 4. Sensor data of two 3D samples computed with k-Wave and s-Wave. (a) 3D view of sample 4. (b) 3D view of sample 5. (c) Sensor data of sample 4 from k-Wave simulation. (d) Sensor data of sample 4 from s-Wave simulation. (e) Sensor data of sample 5 from k-Wave simulation. (f) Sensor data of sample 5 from s-Wave simulation.

IV. CONCLUSION

In this paper, we propose a novel PA simulation method: s-Wave. With easy loop and multiplication operators, we can get the sensor data with much less time cost, especially in sparse system configurations (such as sample with sparse initial pressure distribution in 3D).

In the future work, we will apply our method in PA effect simulation, dataset generation, image reconstruction with model-based algorithms. We believe with more optimization of s-Wave method, it will substitute k-Wave in some specific application scenarios, especially in 3D PA simulation, and become one of the main PA simulation tools in near future.